\documentclass[sigconf]{acmart}
\AtBeginDocument{%
  }

\copyrightyear{2026}
\acmYear{2026}
\setcopyright{cc}
\setcctype{by}
\acmConference[MM '26]{Proceedings of the 34th ACM International Conference on Multimedia}{November 10--14, 2026}{Rio de Janeiro, Brazil}
\acmBooktitle{Proceedings of the 34th ACM International Conference on Multimedia (MM '26), November 10--14, 2026, Rio de Janeiro, Brazil}
\acmDOI{10.1145/3767308.3835677}
\acmISBN{979-8-4007-2213-4/2026/11}

\usepackage{booktabs}
\usepackage{colortbl}
\usepackage{xcolor}
\usepackage{balance}
\usepackage{url}

\begin{document}

\title{ResPCC: A Loss-Resilient Neural Point Cloud Codec over Lossy Networks}

\author{Xueqin Niu}
\authornote{These authors contributed equally to this work.}
\email{starry-night@sjtu.edu.cn}
\affiliation{%
  \institution{Shanghai Jiaotong University}
  \city{Shanghai}
  \country{China}
}

\author{Mufan Liu}
\authornotemark[1]
\email{sudo\_evan@sjtu.edu.cn}
\affiliation{%
  \institution{Shanghai Jiaotong University}
  \city{Shanghai}
  \country{China}
}

\author{Yifan Wang}
\email{yifanw9@sjtu.edu.cn}
\affiliation{%
  \institution{Shanghai Jiaotong University}
  \city{Shanghai}
  \country{China}
}

\author{Le Yang}
\email{le.yang@canterbury.ac.nz}
\affiliation{%
  \institution{University of Canterbury}
  \city{Christchurch}
  \country{New Zealand}
}

\author{Yiling Xu}
\authornote{Corresponding author.}
\email{yl.xu@sjtu.edu.cn}
\affiliation{%
  \institution{Shanghai Jiaotong University}
  \city{Shanghai}
  \country{China}
}

\author{Jun Sun}
\email{junsun@sjtu.edu.cn}
\affiliation{%
  \institution{Shanghai Jiaotong University}
  \city{Shanghai}
  \country{China}
}

\renewcommand{\shortauthors}{Xueqin Niu et al.}

\begin{abstract}
Point cloud compression (PCC) is critical for efficient storage and transmission of 3D data. While recent learning-based PCC methods achieve good rate-distortion (R-D) performance, they generally rely on ideal transmission conditions. In practice, packet loss is a common issue and can severely distort latent features, causing coordinate drift and geometric degradation. To address this challenge, we present ResPCC, the first end-to-end neural point cloud codec designed to offer intrinsic resilience against data loss. Our framework is loss-rate-aware and adapts to diverse packet loss conditions. At the encoder, we introduce a Condition-Adaptive Latent Modulation (CALM) module to adjust latent feature distributions according to the perceived loss rate, as well as a Spatial-Channel Interleaving (SCI) mechanism that transforms channel-wise data extinction into spatially scattered element-wise missing patterns. At the decoder, we develop a Mask-Aware Graph-based Latent Restoration (MGLR) module, followed by a Dictionary-based Refinement (DBR) stage to recover corrupted features and align them with canonical priors. Evaluations on ShapeNet and SemanticKITTI under 5\% to 30\% packet loss rates show that ResPCC consistently delivers superior stability and R-D performance over baselines. Our framework maintains high reconstruction fidelity under lossy conditions, providing a reliable solution for 3D data transmission over practical networks. Code is available at \url{https://github.com/starrynight314/ResPCC}.
\end{abstract}


\maketitle
\vspace{-5mm}
\begin{figure}[h]
  \centering
  \includegraphics[width=0.95\linewidth]{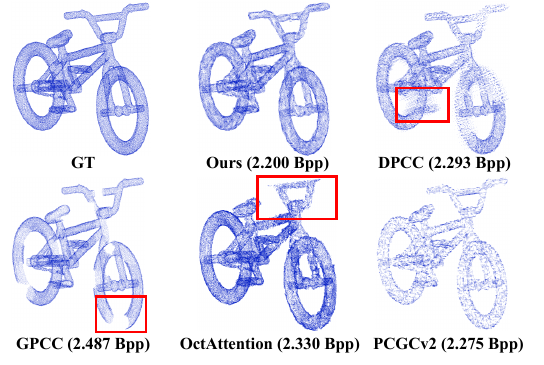}
  \vspace{-3mm}
  \caption{Impact of packet loss on the reconstruction quality of existing codecs.}
  \label{fig:visual_comparison_intro}
\end{figure}
\vspace{-6mm}

\section{INTRODUCTION}
Point clouds have become a fundamental representation for 3D perception and communications \cite{[49],[50],[55]}. However, their substantial data volume often requires efficient compression before storage and transmission. In practical network environments, the resulting bitstreams are packetized and transmitted over lossy networks, making them susceptible to packet loss. When packet loss occurs, the received data may be incomplete, causing the reconstructed point cloud to exhibit geometric distortion and structural degradation. Attaining both compression efficiency and robustness against packet loss is thus essential for reliable point cloud delivery.

Over the years, point cloud compression (PCC) has been studied under traditional and learning-based frameworks. Unlike images or videos which have a fixed grid structure, point clouds are unordered and irregular. As a result, PCC typically begins with organizing the raw geometry into a more structured form. Traditional PCC achieves this using handcrafted techniques, with MPEG G-PCC and V-PCC \cite{[5],[7]} as two representative standards based on hierarchical partitions and projected patches. Learning-based PCC uses a similar compression pipeline, but instead learns compact latent features from hierarchically downsampled point sets and reconstructs the geometry using a neural decoder \cite{[10],[11],[12],[19],[20],[26],[27]}. 

Most existing PCC methods are developed under the assumption that the compressed bitstream is lossless during transmission. Their performance could degrade severely when packet loss occur. This is illustrated in Fig.~\ref{fig:visual_comparison_intro}. Traditional codecs, such as G-PCC, are highly sensitive to bitstream corruption. They may completely fail to decode or exhibit the \emph{cliff effect} \cite{[52]}, where even minor packet loss results in missing regions in the reconstruction. For learning-based codecs, incomplete latent representations due to packet loss tend to disrupt reconstruction more widely, resulting in sparsity artifacts (PCGCv2 \cite{[26]}) or structural distortion (DPCC \cite{[17]}). To mitigate these issues, existing solutions typically rely on either transmission protection or error concealment. Transmission protection schemes, such as forward error correction (FEC) or automatic repeat request (ARQ), can protect the bitstream but introduce additional bandwidth overhead or retransmission delay. Error concealment attempts to infer missing content post-decoding, but lacks reliable priors for faithful restoration when packet loss causes global distortion \cite{[45]}. These strategies implicitly treat robustness as an add-on feature rather than an indispensable part of the codecs for point clouds, leaving the packet-loss-aware design largely unexplored.

This paper presents \emph{ResPCC}, the first end-to-end neural point cloud codec with intrinsic resilience to packet loss for intra-frame geometry compression. ResPCC accounts for packet loss directly in the codec design across encoding, transmission, and decoding. At the encoder, a Condition-Adaptive Latent Modulation (CALM) module conditions latent generation on the packet loss rate information, enabling adaptation to different transmission scenarios. For transmission, a Spatial-Channel Interleaving (SCI) module re-orders the latent representation before packetization. It is capable of spreading the impact of channel-wise dropout into spatially scattered missing patterns, leading to significantly easier content recovery. Packet loss is explicitly simulated on the packetized latent representation during training under both packet-free and packet-loss conditions, allowing the model to achieve high reconstruction fidelity while adapting to lossy transmission conditions. At the decoder, a restoration pipeline consisting of Mask-Aware Graph-based Latent Restoration (MGLR) and Dictionary-based Refinement (DBR) recovers corrupted latent features by aggregating information from neighboring anchors around the packet-loss-affected regions, and further aligns them with canonical priors. We evaluate ResPCC on ShapeNet \cite{[47]} and SemanticKITTI \cite{[48]} across a wide range of packet loss rates. Compared with existing traditional and learning-based PCC methods, ResPCC delivers stronger resilience to transmission loss while maintaining a high reconstruction quality and competitive efficiency. Our main contributions are:
\vspace{-1mm}

\begin{itemize}
    \item We propose ResPCC, the first end-to-end neural point cloud compression codec with intrinsic resilience to packet loss for intra-frame geometry compression;
    \item We adopt a packet-loss-aware codec design that integrates CALM for loss-rate-adaptive latent generation, SCI for corruption dispersion, and a restoration pipeline with MGLR and DBR for recovering corrupted latent representations;
    \item We introduce a training strategy that explicitly simulates both packet-free and packet-loss conditions;
    \item Experiments show that a single ResPCC model generalizes across diverse loss rates while maintaining strong rate-distortion performance and competitive efficiency.
\end{itemize}

\section{RELATED WORKS}
\subsection{Point Cloud Compression}
\textbf{Traditional methods.} Hierarchical tree structures are widely used in traditional PCC. Early works employ KD-trees \cite{[1],[2]} and quadtrees \cite{[4]}, while octree-based methods \cite{[5],[6],[7],[8]} are particularly effective and forms the foundation of the MPEG G-PCC standard \cite{[5],[7]}. G-PCC achieves excellent performance for sparse static point clouds. In contrast, V-PCC \cite{[5],[7]} targets dense or dynamic sequences. Despite its efficiency, G-PCC is highly sensitive to packet loss due to its hierarchical dependencies and state-dependent entropy coding.

\textbf{Learning-based methods. }Deep learning has been widely applied in data compression \cite{[13],[57],[20]} to learn compact representations. Learned PCC can be categorized into point-, voxel-, and octree-based methods \cite{[56]}. \textit{Point-based methods} operate on raw point coordinates, enabling continuous coordinate representation and avoiding quantization artifacts. Existing methods typically adopt hierarchical autoencoder architectures \cite{[10],[11],[12]} and often employ patch-wise strategies \cite{[13],[16],[17],[18]}. Our proposed ResPCC follows this point-based, patch-wise paradigm. \textit{Voxel-based methods} \cite{[19],[20],[24]} map point clouds to volumetric grids, where sparse convolutions \cite{[25]} and tensors \cite{[26],[27]} mitigate computational costs. \textit{Octree-based methods} encode octree occupancy sequences with context-based entropy models. Recent works replace hand-crafted designs with deep models to capture spatial dependencies, including MLP-based \cite{[30],[31]}, 3D CNN-based \cite{[32],[33]}, and Transformer-based approaches \cite{[34],[35],[36]}. Despite significant progress, existing learned PCC methods mainly focus on rate-distortion (R-D) optimization. The impact of packet loss is often overlooked. When packet loss occurs, incomplete latent features are typically zero-padded during decoding. This leads to severe coordinate drift and geometric distortion. The problem becomes more prominent in octree-based models due to their hierarchical and autoregressive dependencies.
\vspace{-2mm}
\subsection{Loss-Resilient Methods}
Various loss-resilient strategies have been developed. Traditional approaches mainly rely on FEC \cite{[38],[39],[41]}, which introduces redundancy to resist loss at the cost of extra bitrate overhead, consuming the budget that could be allocated to improve source coding quality. Recent efforts in the field of image and video coding \cite{[42],[43]} explored intrinsic resilience within learning-based compression. In 3D point cloud compression, however, such intrinsic resilience remains largely unexplored. While some studies \cite{[44],[45],[46]} have investigated error concealment for dynamic point clouds, these methods are dependent on temporal context and are inapplicable to intra-frame compression and often computationally expensive.

\section{PRELIMINARIES}
\subsection{Standard PCC Framework}
A point cloud is defined as a set of points in the 3D space, denoted by $\mathbf{x} = \{\mathbf{x}_i \in \mathbb{R}^3\}_{i=1}^N$. In learning-based PCC, the coding process is typically formulated as a nonlinear transform coding problem. Given an input point cloud $\mathbf{x}$, an analysis transform $g_a(\cdot; \boldsymbol{\theta})$ maps the geometric structure into a latent representation $\mathbf{y} \in \mathbb{R}^{N_S \times C}$:
\begin{equation}
    \mathbf{y} = g_a(\mathbf{x}; \boldsymbol{\theta}),
\end{equation}
where $N_S$ denotes the number of sparse geometric anchors obtained through hierarchical downsampling, and $C$ denotes the dimensionality of the associated latent features. The latent representation $\mathbf{y}$ is then quantized and entropy-coded into a bitstream, and a synthesis transform $g_s(\cdot; \boldsymbol{\phi})$ reconstructs the point cloud as $\hat{\mathbf{x}} = g_s(\mathbf{y}; \boldsymbol{\phi})$. With slight abuse of notations, $\mathbf{y}$ denotes the \textit{quantized} latent representation from now on. PCC minimizes the R-D objective:
\begin{equation}
    \min_{\boldsymbol{\theta}, \boldsymbol{\phi}} 
    \mathcal{L} = D(\mathbf{x}, g_s(\mathbf{y}; \boldsymbol{\phi})) + \lambda R(\mathbf{y}),
\end{equation}
where $D(\cdot,\cdot)$ measures the reconstruction distortion, and $R(\mathbf{y})$ is the bitrate of the latent features.

\subsection{DPCC Backbone}
Following the standard learned PCC practice, we adopt DPCC as our backbone, which is a symmetric autoencoder mapping the input point cloud $\mathbf{x}$ into a decoupled representation through $S$ hierarchical downsampling stages (see Fig.~\ref{fig:framework}, top). At each stage $s \in \{1, \dots, S\}$, Farthest Point Sampling (FPS) identifies $N_s$ sparse anchors. The discarded points are assigned to their nearest anchors so as to form collapsed point sets. The geometric information of these sets is mapped into stage-specific latent features $\mathbf{y}_{\text{ori}}^{(s)} \in \mathbb{R}^{N_s \times C}$. Specifically, this tensor is synthesized by fusing three components via a linear layer, namely the density feature $\mathbf{F}_{\text{dens}}^{(s)}$, the position feature $\mathbf{F}_{\text{pos}}^{(s)}$, and the ancestor feature $\mathbf{F}_{\text{anc}}^{(s)}$, which respectively characterize the local density, relative spatial structure, and hierarchical context around the anchors at stage $s$. After $S$ stages, the encoder outputs the final $N_S$ geometric anchors $\mathbf{A} \in \mathbb{R}^{N_S \times 3}$ and the associated bottleneck latent features $\mathbf{y}_{\text{ori}} \in \mathbb{R}^{N_S \times C}$. The geometric anchors $\mathbf{A}$ are represented in half-precision for transmission, while the latent tensor $\mathbf{y}_{\text{ori}}$ undergoes quantization and entropy coding to generate the compressed bitstream. During decoding, the synthesis transform $g_s(\cdot;\boldsymbol{\phi})$ performs inverse mapping in terms of $S$ symmetric upsampling stages. Guided by $\mathbf{y}_{\text{ori}}$, the decoder predicts the upsampling factor and spatial offset vectors of generated sub-points for each anchor, progressively reconstructing the dense point cloud while preserving reasonable density distributions.

However, like most existing learned PCC frameworks, this standard pipeline is developed under the implicit assumption of reliable transmission and does not account for packet loss during latent delivery. In practical lossy networks, missing latent data packets can induce a distribution shift between the transmitted and received representations, which may further cause manifold deviation and topological collapse during decoding, ultimately leading to severe geometric distortion. This motivates the study of packet-loss-aware learning-based PCC under realistic lossy transmission settings.

\section{PROBLEM FORMULATION}
We shall formulate the lossy PCC task as a packet-loss-aware R-D optimization problem. It is assumed that over a short transmission interval, the packet loss condition is approximately fixed and can be described using a proxy perceived loss rate $p$. With this assumption, given an input point cloud $\mathbf{x}$, the encoder produces a latent representation $\mathbf{y}=g_a(\mathbf{x}, p; \boldsymbol{\theta})$ that is explicitly dependent on $p$. During transmission, $\mathbf{y}$ is partitioned into $K$ disjoint packets $\{\mathbf{y}^{(k)}\}_{k=1}^K$. Let $\mathbf{M} \in \{0,1\}^K$ denote the packet reception indicator, where $M_k=1$ indicates the successful reception of the $k$-th packet and $M_k=0$ indicates packet loss. The corrupted latent representation received by the decoder can be denoted as
\begin{equation}
    \tilde{\mathbf{y}}^{(k)} = M_k \mathbf{y}^{(k)}, \quad \forall k \in \{1,\dots,K\}.
\end{equation}

Stacking $\tilde{\mathbf{y}}^{(k)}$ yields the corrupted latent code $\tilde{\mathbf{y}}$ under packet loss pattern $\mathbf{M}$. To recover missing information before synthesis, we introduce a restoration operator $h(\cdot;\boldsymbol{\psi})$ acting on $\tilde{\mathbf{y}}$. The robust packet-loss-aware PCC task is then formulated as
\begin{equation}
    \min_{\boldsymbol{\theta}, \boldsymbol{\phi}, \boldsymbol{\psi}}
    \mathcal{J}
    =
    \mathbb{E}_{\mathbf{M}\sim\mathbb{P}(\mathbf{M}\mid p)}
    \left[
    D\!\left(
    \mathbf{x},
    g_s\!\left(h(\tilde{\mathbf{y}};\boldsymbol{\psi});\boldsymbol{\phi}\right)
    \right)
    \right]
    + \lambda R(\mathbf{y}).
    \label{RD}
\end{equation}

\section{PROPOSED METHOD}
\subsection{Framework Overview}
\begin{figure*}[t] 
  \centering
  \includegraphics[width=\textwidth]{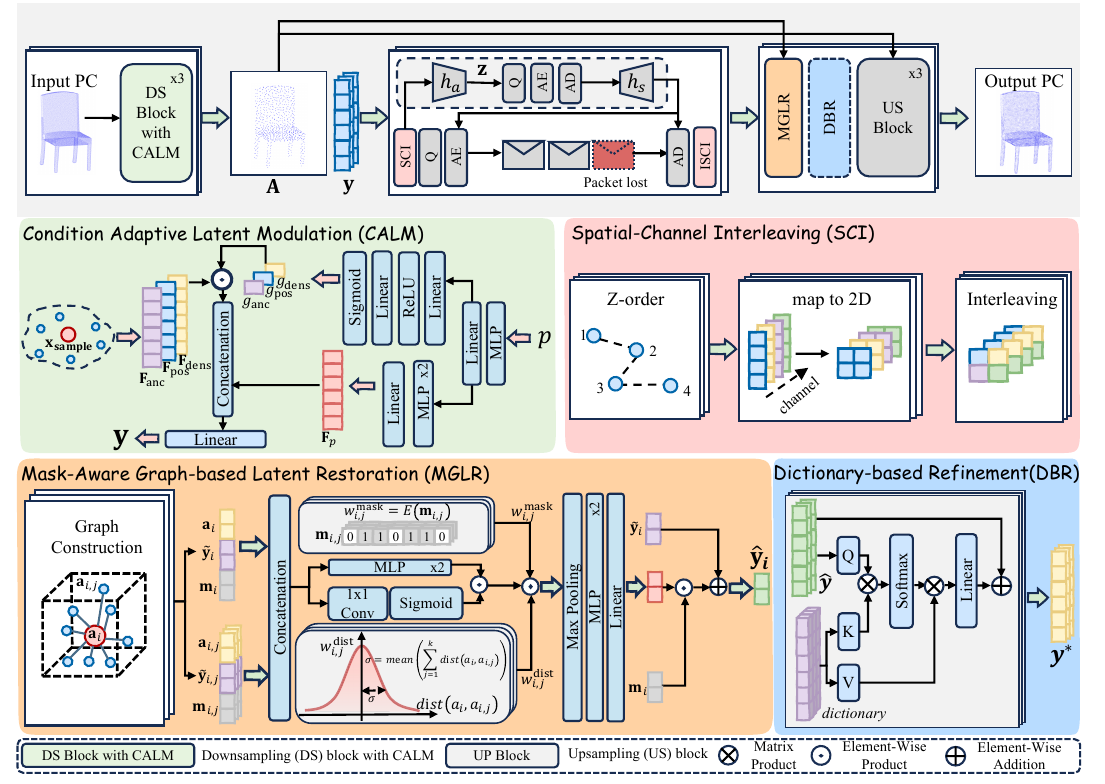} 
  \caption{The overall framework of ResPCC. The encoder generates anchors $\mathbf{A}$ and latent features $\mathbf{y}$, where CALM adjusts the feature distribution based on loss rates. The SCI module rearranges the latent representations to transform potential channel-wise loss into scattered element-wise missing values. At the decoder, MGLR restores corrupted features $\mathbf{\tilde{y}}$ using spatial context from $\mathbf{A}$ and the loss mask $\mathbf{m}$, followed by DBR and final upsampling. Sub-modules are detailed below.}
  \Description{overview}
  \label{fig:framework}
\end{figure*}

The overall pipeline of ResPCC is illustrated in Fig.~\ref{fig:framework}. Built upon the DPCC backbone, ResPCC offers improved robustness to packet loss by introducing CALM at the encoder, SCI before packetization, and MGLR with DBR at the decoder.

At the encoder, the input point cloud is progressively downsampled to produce geometric anchors and latent features. CALM modulates intermediate features according to the estimated loss rate for loss-aware encoding. The anchors, which occupy only a small fraction of the bitstream (e.g., $\sim$0.38 bpp) but serve as the structural skeleton for subsequent upsampling, are treated as metadata and transmitted reliably with FEC, where the redundancy overhead is negligible due to the low bitrate used. The latent features are then re-arranged by SCI across spatial and channel dimensions before packetization, converting packet-loss-induced disruption into scattered or even isolated missing values. The re-arranged features are first processed by a variational hyperprior \cite{[51]} to estimate their probability distribution for compression, and then quantized and partitioned into packets. Each packet is independently entropy-coded using the estimated distributions. The corresponding hyperprior information, accounting for only 0.4\%--0.7\% of the total bitrate, is also transmitted via FEC as protected metadata to ensure entropy model consistency. At the decoder, the corrupted latent representation is first passed through inverse SCI, then restored by MGLR using correlations among neighboring anchors, and finally refined by the DBR to better align with canonical latent priors. This design enables robust and high-fidelity point cloud reconstruction under lossy transmissions.

\subsection{Condition-Adaptive Latent Modulation}
Training with simulated packet loss \cite{[42]} equips the codec with basic robustness. However, without explicit conditioning, training across a wide range of loss conditions yields suboptimal and non-adaptive latent features, leading to inconsistent performance under varying loss rates. In contrast, methods trained under a fixed loss condition require multiple models to handle different loss levels, making them impractical in real-world scenarios. To address this issue, we propose the CALM module, which leverages the estimated packet loss rate $p$ to dynamically regulate the encoding process.

Given the estimated packet loss rate $p$, each CALM module at stage $s \in \{1, \dots, S\}$ independently maps $p$ into a high-dimensional condition embedding through an MLP with frequency-aware encoding. Although the input condition $p$ is shared across all stages, the CALM modules do not share parameters, resulting in stage-specific embeddings and modulation behaviors. The resulting embedding is then fed into two parallel branches. One branch generates a packet-loss-aware feature $\mathbf{F}_p^{(s)} \in \mathbb{R}^{C}$, while the other produces three gating factors $\mathbf{G}^{(s)} = \{g_{\text{anc}}^{(s)}, g_{\text{pos}}^{(s)}, g_{\text{dens}}^{(s)}\}$ to adjust the relative importance of the ancestor, position, and density features, which are denoted by $\mathbf{F}_{\text{anc}}^{(s)}$, $\mathbf{F}_{\text{pos}}^{(s)}$, and $\mathbf{F}_{\text{dens}}^{(s)}$, respectively. At the $s$-th hierarchical stage with $N_s$ sampled anchors, the three geometric features are first modulated using
\begin{equation}
\tilde{\mathbf{F}}_{\text{anc}}^{(s)} = g_{\text{anc}}^{(s)} \mathbf{F}_{\text{anc}}^{(s)}, \quad
\tilde{\mathbf{F}}_{\text{pos}}^{(s)} = g_{\text{pos}}^{(s)} \mathbf{F}_{\text{pos}}^{(s)}, \quad
\tilde{\mathbf{F}}_{\text{dens}}^{(s)} = g_{\text{dens}}^{(s)} \mathbf{F}_{\text{dens}}^{(s)}.
\end{equation}

The feature $\mathbf{F}_p^{(s)}$ is shared along the spatial dimension and fused with the modulated geometric features through channel-wise concatenation. Applying an extra linear projection to the result gives
\begin{equation}
\mathbf{y}^{(s)} =
\mathrm{Linear}\!\left(
\left[
\tilde{\mathbf{F}}_{\text{anc}}^{(s)},
\tilde{\mathbf{F}}_{\text{pos}}^{(s)},
\tilde{\mathbf{F}}_{\text{dens}}^{(s)},
\mathbf{F}_p^{(s)}
\right]
\right),
\end{equation}
where $[\cdot]$ denotes channel-wise concatenation. In this way, CALM injects the packet loss prior into each hierarchical stage through stage-specific conditioning, enabling the encoder to produce latent features $\mathbf{y}^{(s)} \in \mathbb{R}^{N_s \times C}$ that are adaptively modified according to the current packet loss rate.

\vspace{-2mm}
\subsection{Spatial-Channel Interleaving}
\begin{figure}[t]
  \centering
  \includegraphics[width=\linewidth]{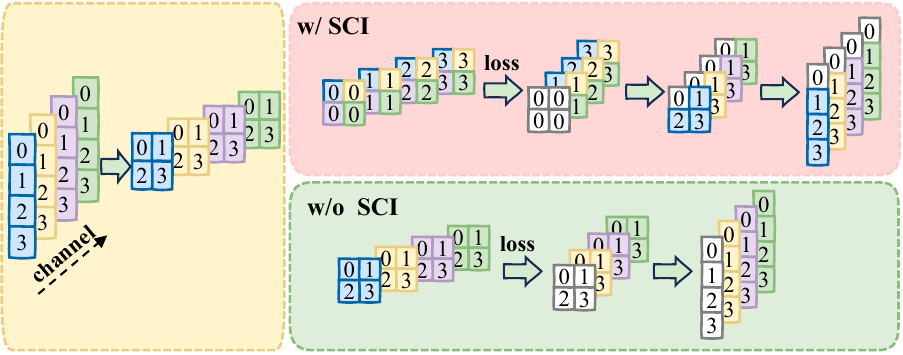}
  \caption{Rearranging $\mathbf{y}_{2D}$ via SCI. White elements denote lost data. While the absence of SCI results in channel-wise data extinction, SCI transforms such damage into element-wise missing values distributed across multiple channels.}
  \vspace{-4mm}
  \label{fig:sci_mechanism}
\end{figure}

As illustrated in Fig.~\ref{fig:sci_mechanism}, with channel-wise packetization, the loss of one packet can result in the removal of the relative feature dimensions for all anchors, leaving insufficient references and making subsequent feature restoration more challenging. To address this, we propose the SCI mechanism and its inverse transform (ISCI) to spread the packet-loss-induced damage in the latent space. Given geometric anchors $\mathbf{A}\in\mathbb{R}^{N_S\times 3}$ and their latent features $\mathbf{y}\in\mathbb{R}^{N_S\times C}$ from the final downsampling stage, we first compute a permutation $\pi$ by sorting the 3D points in $\mathbf{A}$ according to their Morton order, and re-organize both anchors and latent features accordingly using
\begin{equation}
\mathbf{A}'=\pi(\mathbf{A}), \qquad \mathbf{y}'=\pi(\mathbf{y}).
\end{equation}

\vspace{-3mm}
As a result, spatially nearby anchors now become adjacent. We then reshape $\mathbf{y}'$ into a padded 2D latent tensor $\mathbf{y}_{2D}\in\mathbb{R}^{H\times W\times C}$ with $HW\ge N_S$, where $(H,W)$ is selected from a predefined lookup table according to $N_S$ to minimize the required number of zero padding while remaining compatible with the hyperprior network. SCI is next applied to $\mathbf{y}_{2D}$ over each local $2\times2\times4$ block, yielding an interleaved tensor
\begin{equation}
\mathbf{y}_{2D}^{i}=\mathrm{SCI}(\mathbf{y}_{2D}).
\end{equation}

More specifically, for each spatial block indexed by $(i,j)$ and each four-channel group indexed by $t$, let
\begin{equation}
\mathcal{B}_{i,j,t}=\mathbf{y}_{2D}[2i:2i+2,\;2j:2j+2,\;4t:4t+4]\in\mathbb{R}^{2\times2\times4},
\end{equation}
We further define the elements inside each block using the triplet $(u,v,k)$, where $u,v \in \{0,1\}$ and $k \in \{0,1,2,3\}$. SCI applies a fixed bijective permutation so that the four elements from each original channel are redistributed to different spatial positions and channels. In particular, the output tensor $\mathbf{y}_{2D}^{i}$ is
\begin{equation}
\mathbf{y}_{2D}^{i}(2i+u',\; 2j+v',\; 4t+k') 
= \mathbf{y}_{2D}(2i+u,\; 2j+v,\; 4t+k),
\end{equation}
where the index mapping is defined as
\begin{equation}
    u' = k \bmod 2, \quad v' = \lfloor k/2 \rfloor, \quad k' = 2u + v.
\end{equation}

During transmission, let $\tilde{\mathbf{y}}_{2D}^{i}$ denote the corrupted received tensor and let $\mathbf{m}^{i}\in\{0,1\}^{H\times W\times C}$ be the corresponding binary mask indicating valid entries. The decoder then applies the inverse transform ISCI to recover the original spatial arrangement via
\begin{equation}
(\tilde{\mathbf{y}}_{2D},\mathbf{m}_{2D})=\mathrm{ISCI}(\tilde{\mathbf{y}}_{2D}^{i},\mathbf{m}^{i}),
\end{equation} 
Finally, $\tilde{\mathbf{y}}_{2D}$ is reshaped back to the serialized latent sequence $\tilde{\mathbf{y}}\in\mathbb{R}^{N_S\times C}$ together with its mask $\mathbf{m}$. In this way, SCI converts the packet-loss-induced channel-wise data extinction into spatially scattered missing patterns, while retaining local references for the subsequent restoration.

\subsection{Mask-Aware Restoration and Refinement}

At the decoder, the corrupted latent features $\tilde{\mathbf{y}}\in\mathbb{R}^{N_S\times C}$ together with its binary mask $\mathbf{m}\in\{0,1\}^{N_S\times C}$ indicating the missing entries due to packet loss are received. The restoration objective is to recover the corrupted entries while preserving the correctly received ones. To this end, we first perform mask-aware graph-based latent restoration by exploiting local geometric and feature correlations among neighboring anchors, and then optionally invoking a lightweight dictionary-based refinement module to further align the restored features with canonical latent priors.

\textbf{Mask-Aware Graph-based Latent Restoration.}
We construct a $K$-nearest-neighbor graph $\mathcal{G}$ over the anchor coordinates $\mathbf{A}=\{\mathbf{a}_i\}_{i=1}^{N_S}$, where $\mathcal{N}(i)$ denotes the neighborhood of anchor $i$. For each pair $(i,j)$ with $j\in\mathcal{N}(i)$, an edge descriptor is formed as
\begin{equation}
\mathbf{e}_{ij}=\mathrm{MLP}\!\left([\mathbf{a}_i,\mathbf{a}_j,\tilde{\mathbf{y}}_i,\tilde{\mathbf{y}}_j,\mathbf{m}_i,\mathbf{m}_j]\right),
\end{equation}
where $[\cdot]$ denotes channel-wise concatenation. To favor reliable and spatially relevant neighbors, we define a mask-aware reliability weight and a distance-aware weight as
\begin{equation}
w_{ij}^{\mathrm{mask}}=\frac{1}{C}\sum_{c=1}^{C} m_{j,c}, \qquad
w_{ij}^{\mathrm{dist}}=\exp\!\left(-\frac{\|\mathbf{a}_i-\mathbf{a}_j\|_2^2}{2\sigma_i^2}\right),
\end{equation}
where $m_{j,c}$ denotes the $c$-th mask entry of neighbor $j$, and $\sigma_i$ is an adaptive local scale estimated from the distances within $\mathcal{N}(i)$. The weighted edge feature is then computed by the following element-wise multiplication
\begin{equation}
\tilde{\mathbf{e}}_{ij}=\mathbf{e}_{ij}\odot w_{ij}^{\mathrm{mask}} \odot w_{ij}^{\mathrm{dist}},
\end{equation}
The restoration term is obtained by performing element-wise max pooling:
\begin{equation}
\mathbf{\delta}_i = \max_{j \in \mathcal{N}(i)} (\tilde{\mathbf{e}}_{ij}),
\end{equation}
where $\max(\cdot)$ denotes the element-wise maximum operator.
The latent features are finally updated using the masked residual compensation
\begin{equation}
\hat{\mathbf{y}}=\tilde{\mathbf{y}}+(\mathbf{1}-\mathbf{m})\odot\boldsymbol{\delta},
\end{equation}
Note that only corrupted entries are corrected while valid entries remain intact.

\begin{figure*}[h]
  \centering
  \includegraphics[width=\textwidth]{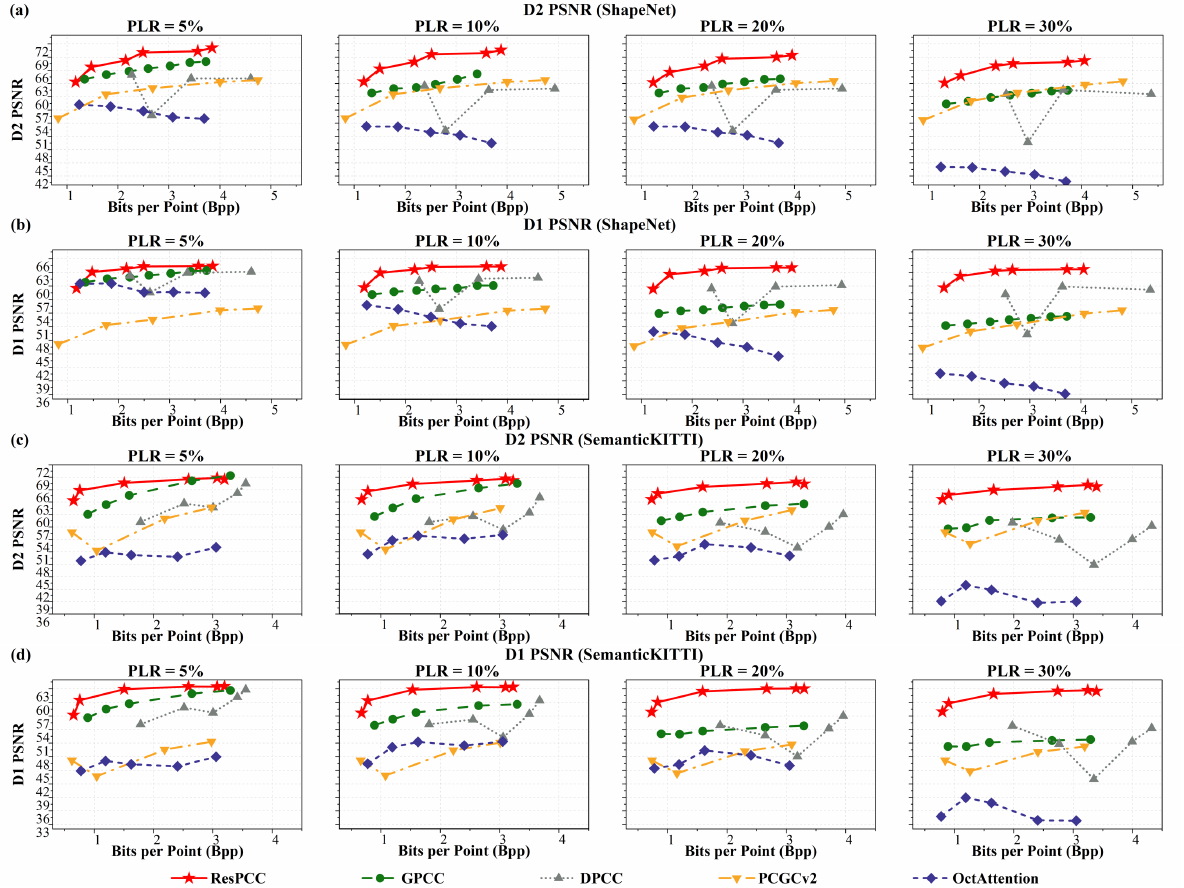} 
  \caption{Comparison of R-D performance between ResPCC and baselines under packet loss rates (PLR) from 5\% to 30\%. (a)-(b) present D2 and D1 PSNR on the ShapeNet dataset. (c)-(d) summarize results on the SemanticKITTI dataset.}
  \label{fig:rd_curves}
\end{figure*}

\textbf{Dictionary-based Refinement.}
Although the above graph-based restoration recovers most missing information, it may introduce over-smoothing in the latent space. To alleviate this, we optionally introduce a learnable codebook $\mathbf{D}\in\mathbb{R}^{M\times C}$ consisting of $M$ prototype latent patterns. The restored features $\hat{\mathbf{y}}$ are used as queries in a cross-attention operation over $\mathbf{D}$, and the retrieved prior features are added back with a learnable scaling factor to obtain the final refined representation:
\begin{equation}
\mathbf{y}^{*} = \hat{\mathbf{y}} + \gamma \, \mathrm{Attn}(\hat{\mathbf{y}}, \mathbf{D}, \mathbf{D}),
\end{equation}
where $\mathbf{y}^{*}$ denotes the final refined latent representation. This refinement step regularizes the restored features utilizing canonical latent priors and improves the quality of subsequent point cloud reconstruction.

\subsection{Simulated Packet Loss}
To incorporate both packet-free and packet-loss conditions into end-to-end training, we simulate lossy transmission on the interleaved latent representation. The packet loss rate $p$ is sampled from a mixed distribution: with a certain probability, $p=0$; otherwise, it is drawn from a Beta distribution and rescaled to a predefined range. Accordingly, when $p=0$, the binary mask is set to $\mathbf{m}^{i}=\mathbf{1}$, corresponding to packet-free transmission; otherwise, $\mathbf{m}^{i}\in\{0,1\}^{H\times W\times C}$ is sampled according to $p$, yielding the corrupted latent tensor
\begin{equation}
\tilde{\mathbf{y}}_{2D}^{i}=\mathbf{m}^{i}\odot \mathbf{y}_{2D}^{i}.
\end{equation}

\section{PERFORMANCE EVALUATION}
\subsection{Experiment Setup}

\textbf{Datasets. }We use ShapeNet \cite{[47]} and SemanticKITTI \cite{[48]}. Following \cite{[17]}, we scale global coordinates to $128^3$ (ShapeNet) and $100^3$ (SemanticKITTI), then partition them into non-overlapping patches of $32^3$ and $12^3$, respectively. Finally, we normalize each patch's coordinates to $[-1, 1]$ for encoding and decoding.

\begin{figure*}[h]
  \centering
  \includegraphics[width=\textwidth]{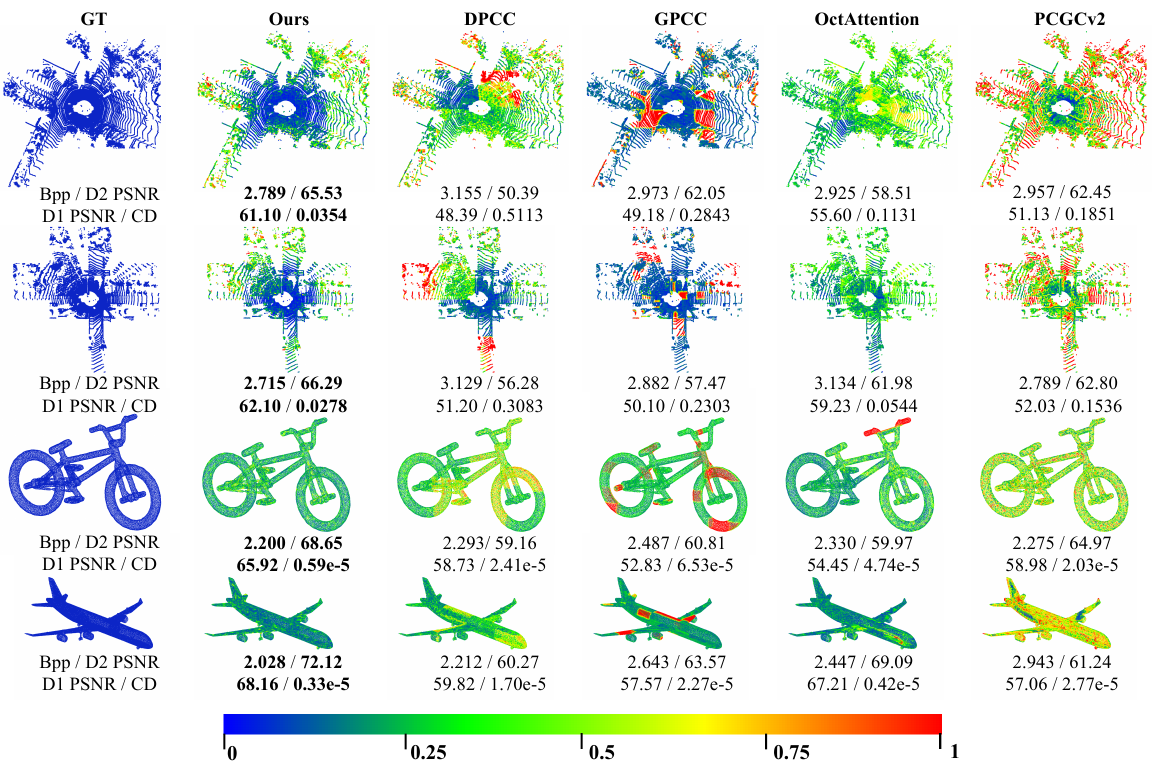} 
  \caption{Qualitative comparison under 20\% packet loss. The color bar indicates reconstruction error, where the maximum value (1.0) corresponds to 1\,mm for ShapeNet and 1\,m for SemanticKITTI. While baselines suffer from significant structural loss or distortion, ResPCC maintains superior geometric integrity and lower error across all scenes.}
  \label{fig:visual}
\end{figure*}

\textbf{Training Settings. }
The framework is built upon a structurally optimized DPCC baseline. The downsampling rate is increased from $1/3$ to $1/5$ per stage, reducing the anchor bitrate from 1.78 bpp to 0.38 bpp. The latent channel dimension is expanded from 8 to 64, and deeper convolutional layers are adopted. During training, the loss rate is set to zero with the probability of 50\%; otherwise, it is sampled from a Beta distribution ($\alpha=1.1, \beta=2$) and then scaled to a $[0, 50\%]$ packet loss rate interval. We use the Adam optimizer with an initial learning rate of $10^{-3}$, and train the model on an NVIDIA RTX 3090 GPU. The backbone is first trained for 50 epochs, followed by 15 epochs of fine-tuning the dictionary and decoder with the backbone frozen.

\textbf{Testing Settings. } To simulate realistic bursty packet loss, we employ the Gilbert-Elliott (G-E) model \cite{[53]} to generate packet loss sequences at target rates of $\{5\%, 10\%, 20\%, 30\%\}$.

\textbf{Evaluation Metrics. } We evaluate performance using D1-PSNR, D2-PSNR, and Bits Per Point (Bpp) \cite{[54]}. The normals required for D2 calculation are obtained from the ShapeNet metadata or estimated via Open3D for SemanticKITTI. For PSNR, the peak value is defined as the diagonal length of each dataset's maximum spatial range.

\textbf{Baselines. } We compare ResPCC with G-PCC (TMC13 v23.0) and three representative paradigms: DPCC \cite{[17]} (point-based), PCGCv2 \cite{[26]} (voxel-based), and OctAttention \cite{[34]} (octree-based). To ensure a fair evaluation under lossy conditions, we adopt method-specific strategies to maintain decodability of all baselines without altering their core encoding mechanisms. For G-PCC, the point cloud is partitioned into slices, ensuring that packet loss only affects localized regions. For PCGCv2 and DPCC, FEC is applied to the anchor bitstreams after the final downsampling stage, and the resulting overhead is included in the Bpp calculations. For OctAttention, we employ an error-tolerant decoding strategy that detects inconsistencies between the octree structure and the available bitstream, and terminates decoding early to avoid failure but still returning partially reconstructed points.

\definecolor{bestblue}{RGB}{232, 244, 255}
\begin{table*}[t]
  \centering
  \caption{Ablation study of ResPCC on the ShapeNet dataset across different PLRs. We evaluate the contribution of each high-level module and the impact of MGLR and SCI. Best results within each category are in \textbf{bold}, and \underline{underline} compares the impact of SCI on the vanilla baseline.}
  \label{tab:commands}
  \vspace{-3mm}
  \resizebox{\textwidth}{!}{
  
  \begin{tabular}{ccccc|ccc|ccc|ccc|ccc}
    \toprule
    \multicolumn{5}{c|}{\textbf{Module Configuration}} & \multicolumn{3}{c|}{\textbf{PLR = 5\%}} & \multicolumn{3}{c|}{\textbf{PLR = 10\%}} & \multicolumn{3}{c|}{\textbf{PLR = 20\%}} & \multicolumn{3}{c}{\textbf{PLR = 30\%}} \\
    DBR & CALM & MGLR & Linear & SCI & Bpp$\downarrow$ & D2$\uparrow$ & D1$\uparrow$ & Bpp$\downarrow$ & D2$\uparrow$ & D1$\uparrow$ & Bpp$\downarrow$ & D2$\uparrow$ & D1$\uparrow$ & Bpp$\downarrow$ & D2$\uparrow$ & D1$\uparrow$ \\
    \midrule

    \multicolumn{17}{l}{\cellcolor{gray!10}\textit{Category I: Effectiveness of CALM and DBR}} \\
    \rowcolor{bestblue}
    $\checkmark$ & $\checkmark$ & $\checkmark$ & $\times$ & $\checkmark$ & \textbf{2.500} & \textbf{70.093} & \textbf{64.337} & \textbf{2.528} & \textbf{69.674} & \textbf{64.230} & \textbf{2.590} & \textbf{68.713} & \textbf{63.972} & \textbf{2.667} & \textbf{67.593} & \textbf{63.595} \\
    $\times$ & $\checkmark$ & $\checkmark$ & $\times$ & $\checkmark$ & 2.500 & 70.012 & 64.317 & 2.528 & 69.598 & 64.211 & 2.590 & 68.642 & 63.949 & 2.667 & 67.526 & 63.569 \\
    $\times$ & $\times$ & $\checkmark$ & $\times$ & $\checkmark$ & 2.653 & 69.152 & 64.112 & 2.678 & 68.697 & 63.990 & 2.736 & 67.670 & 63.669 & 2.811 & 66.530 & 63.210 \\

    \midrule
    \multicolumn{17}{l}{\cellcolor{gray!10}\textit{Category II: Impact of Restoration Backbone and SCI}} \\
    $\times$ & $\times$ & $\checkmark$ & $\times$ & $\checkmark$ & \textbf{2.325} & \textbf{68.850} & \textbf{63.887} & \textbf{2.349} & \textbf{68.196} & \textbf{63.744} & \textbf{2.407} & \textbf{67.218} & \textbf{63.370} & \textbf{2.482} & \textbf{66.062} & \textbf{62.835} \\
    $\times$ & $\times$ & $\times$ & $\checkmark$ & $\checkmark$ & 2.395 & 68.331 & 63.768 & 2.419 & 67.491 & 63.511 & 2.478 & 65.987 & 62.904 & 2.552 & 64.668 & 62.207 \\
    $\times$ & $\times$ & $\times$ & $\times$ & $\checkmark$ & \underline{2.430} & 68.518 & \underline{63.792} & \underline{2.455} & 67.866 & \underline{63.599} & \underline{2.513} & \underline{66.571} & \underline{63.131} & \underline{2.588} & \underline{65.285} & \underline{62.510} \\
    $\times$ & $\times$ & $\times$ & $\times$ & $\times$ & 2.559 & \underline{68.761} & 63.684 & 2.584 & \underline{68.000} & 63.496 & 2.642 & 66.504 & 63.019 & 2.716 & 65.013 & 62.345 \\

    \bottomrule
  \end{tabular}
  }
\end{table*}

\subsection{Results}
\textbf{Evaluation and Comparison with Baseline Methods.} As shown in Fig.~\ref{fig:rd_curves}, ResPCC consistently outperforms all baselines across varying packet loss rates. Under a mild packet loss rate (PLR) of 5\%, it already achieves noticeable gains over DPCC on ShapeNet, with average improvements of 5.10\% in D1 and 8.84\% in D2. At a comparable bitrate (2.48 bpp), ResPCC reaches 70.06 dB in D2-PSNR, surpassing both G-PCC and DPCC, demonstrating strong reconstruction capability even under stable transmission conditions. As the packet loss rate increases, the performance advantage of ResPCC becomes more pronounced. At PLR = 30\%, the average gains over DPCC increase to 14.64\% (D1) and 19.31\% (D2). Notably, compared with OctAttention, ResPCC achieves substantial improvements exceeding 50\% in both D1 and D2, highlighting its robustness to severe packet loss. Even at similar or lower bitrates, ResPCC maintains superiority over all baselines. Similar trends are observed on the SemanticKITTI dataset, where the performance gains consistently enlarge with increasing loss rates. In particular, the gains over G-PCC grow from 4.78\% (D1) and 4.35\% (D2) at 5\% PLR to 20.66\% and 13.23\% at 30\% PLR. Meanwhile, the improvement over DPCC also increases significantly, reaching over 23\% at high loss rates. At PLR = 30\%, ResPCC achieves 67.16 dB (D2) at 3.24 bpp, clearly outperforming PCGCv2 at a comparable bitrate.

The inferior performance of baselines stems from their rigid architectural dependence on data integrity. DPCC suffers from severe coordinate drift, and G-PCC exhibits blocky geometric holes caused by its spatial slicing strategy. OctAttention is particularly unstable, as its reconstruction quality depends heavily on the octree depth at which loss occurs. PCGCv2 exhibits geometric distortions when the loss of packets compromises feature representations. We also observe that these baselines exhibit performance instability where increasing the bitrate may not always yield better results. This suggests that the high-frequency features introduced at higher bitrates might be more sensitive to packet loss.

\textbf{Complexity and Model Size.} We evaluate the storage requirements by considering the size of the binary executable for G-PCC and the checkpoint sizes for other methods. 
\definecolor{bestblue}{RGB}{232, 244, 255}
\vspace{-3mm}
\begin{table}[h]
  \centering
  \caption{Complexity analysis (ShapeNet / SemanticKITTI) and size. Encoding and decoding times (ms) of per-patch measured on an NVIDIA RTX 3090 GPU.}
  \label{tab:complexity}
  \vspace{-2mm}
  \begin{tabular}{lccc}
    \toprule
    \textbf{Methods} & \textbf{Enc. (ms)} & \textbf{Dec. (ms)} & \textbf{Size (MB)} \\
    \midrule
    DPCC        & 23.11 / 78.93      & 12.81 / 17.32      & 1.31  \\
    G-PCC       & 85.32 / 93.62      & 82.17 / 94.56      & 5.25  \\
    PCGCv2      & 391.0 / 179.0      & 376.0 / 194.0      & 3.19  \\
    OctAttention & 140.0 / 136.0      & 9716.6 / 6134.0    & 27.97 \\
    \rowcolor{bestblue}
    \textbf{Ours}    & \textbf{56.11 / 85.17} & \textbf{49.93 / 52.88} & \textbf{21.08} \\
    \bottomrule
  \end{tabular}
\end{table}
\vspace{-3mm}

As shown in Table~\ref{tab:complexity}, while the integration of additional resilient modules leads to increased computational overhead and model size, the complexity remains within an acceptable range for practical deployment. 

\subsection{Ablation Study}
We conduct an ablation study on the ShapeNet dataset to verify the contribution of the proposed modules. To ensure a fair comparison, every configuration is re-trained in an end-to-end manner. 

\textbf{Effectiveness of CALM and DBR.} As illustrated in Table~\ref{tab:commands}, we analyze performance gains by sequentially removing modules from the full ResPCC. Excluding the DBR module consistently degrades reconstruction quality across all PLRs. While D1-PSNR decreases modestly, D2-PSNR drops significantly. As D2-PSNR measures point-to-plane distances, it better represents surface morphology and geometric authenticity. This trend confirms that dictionary priors effectively refine MGLR features through detailed alignment. Furthermore, removing the CALM module further reduces R-D efficiency, yielding lower PSNR even at higher bitrates. This demonstrates that CALM adaptively modulates latent features based on packet loss risk to prioritize critical geometric information.

\begin{figure}[h] 
  \centering
  \includegraphics[width=\columnwidth]{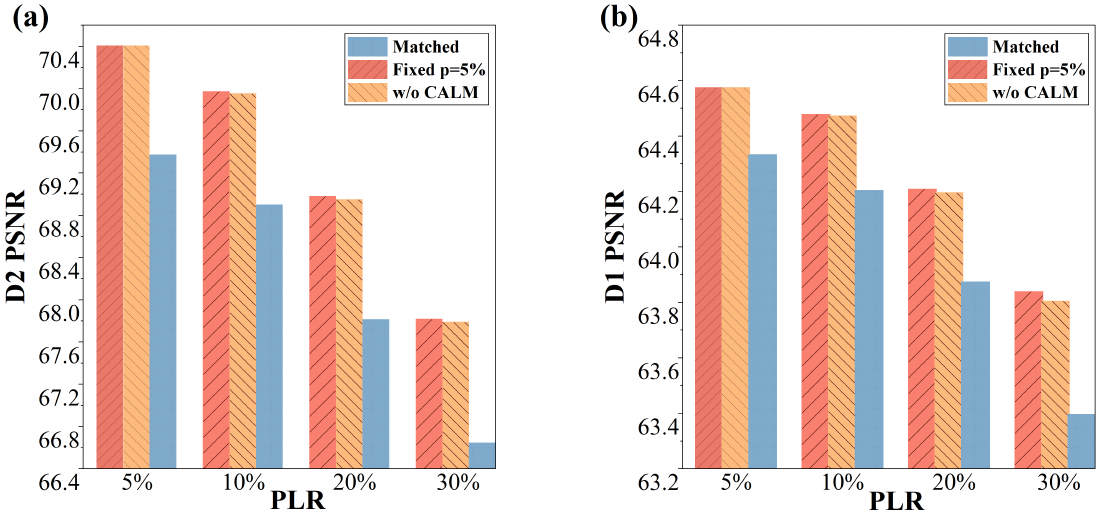} 
  \vspace{-6mm}
  \caption{Sensitivity of ResPCC to prior $p$ on ShapeNet: matched $p$ vs. fixed prior ($p=5\%$) and w/o CALM (all variants are evaluated without DBR module).}
  \label{fig:sensitivity}
  \vspace{-6mm}
\end{figure}

\textbf{Impact of Restoration Backbone and SCI.} As presented in Table~\ref{tab:commands}, we evaluate the effectiveness of the MGLR and SCI modules. To validate MGLR, we implement a linear baseline using the KNN-based spatial interpolation. Results show that MGLR significantly outperforms this baseline; at 30\% PLR, MGLR achieves 1.4 dB higher D2-PSNR at a lower bitrate (2.482 bpp). This confirms that MGLR learns deep local geometric constraints instead of simple numerical interpolation. Furthermore, the \underline{underlined} results highlight SCI's independent impact. Even without a restoration backbone, SCI improves D1/D2-PSNR by scattering channel-wise losses into element-wise missing values. This advantage becomes more pronounced as the packet loss rate increases, as expected.

\textbf{Robustness to Prior Mismatch.} As illustrated in Fig. ~\ref{fig:sensitivity}, we evaluate the robustness of ResPCC against inaccurate loss rate estimations. Even with mismatched priors, ResPCC exhibits minimal degradation and remains superior to the baseline without CALM, making our framework highly attractive for real-world deployment.

\section{CONCLUSIONS}
The paper presented \textbf{ResPCC}, a loss-resilient neural point cloud codec. Through the synergy of \textbf{CALM} for adaptive feature modulation, \textbf{SCI} for packet loss-induced error dispersion, and a dual-stage restoration pipeline (\textbf{MGLR} and \textbf{DBR}), ResPCC achieved high-fidelity reconstruction capability under packet loss. Evaluations on ShapeNet and SemanticKITTI demonstrated that ResPCC maintains superior R-D performance and consistent stability across various loss rates, significantly outperforming existing codecs. With an acceptable increase in processing latency and complexity, ResPCC provides a practically reliable solution to 3D point cloud transmission over lossy networks.

\begin{acks}
This paper is supported in part by National Key R\&D Program of China (2024YFB2907204), National Natural Science Foundation of China (62371290), the Fundamental Research Funds for the Central Universities of China, and STCSM under Grant (22DZ2229005). The corresponding author is Yiling Xu (e-mail: yl.xu@sjtu.edu.cn).
\end{acks}

\bibliographystyle{ACM-Reference-Format}
\balance
\bibliography{sample-base}

\appendix

\section{Detailed Network Architecture}

This section provides a comprehensive description of the ResPCC architecture, including specific layer configurations and tensor shape transformations. All convolutional layers are followed by Group Normalization (GN) and ReLU activation unless specified otherwise.

\subsection{Encoder}
\begin{figure*}[h]
  \centering
  \includegraphics[width=\textwidth]{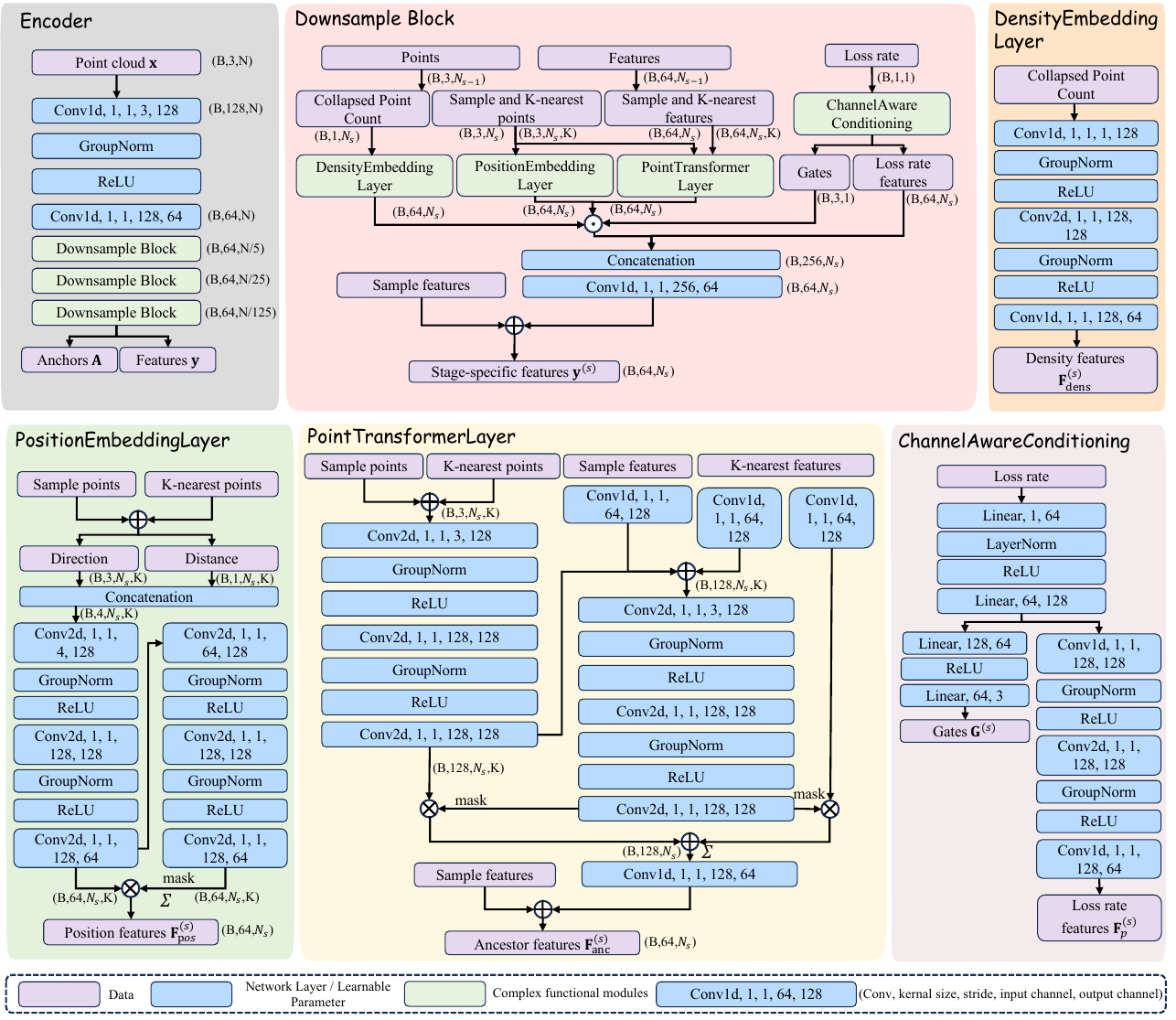} 
  \caption{Detailed architecture of the Encoder and Downsample Layer. The encoder hierarchically extracts features while the CALM module adaptively modulates the feature distribution using gating factors $\mathbf{G}^{(s)}$ and $\mathbf{F}_p^{(s)}$.}
  \label{fig:encoder_ds}
\end{figure*}

The detailed structure of the encoder is illustrated in Figure~\ref{fig:encoder_ds}. The encoder begins with an initial feature embedding stage where the input point cloud $\mathbf{x} \in \mathbb{R}^{B \times 3 \times N}$ is transformed into a 64-channel latent space. This process is executed by two Conv1d layers with a kernel size of 1, shifting the channel dimension from $3 \to 128 \to 64$.

The core of the downsampling hierarchy consists of three stages, where the point count is progressively reduced from $N \to N/5 \to N/25 \to N/125$ via Farthest Point Sampling (FPS). As shown in the Downsample Layer of Figure~\ref{fig:encoder_ds}, the block integrates four parallel branches. The Channel-Aware Conditioning submodule takes the perceived loss rate $p \in \mathbb{R}^{B \times 1 \times 1}$ as input and utilizes a frequency-aware MLP to produce a condition embedding. This embedding is branched into two functional components: 1) a gating generator that produces three scalar factors $\mathbf{G}^{(s)} = \{g_{anc}^{(s)}, g_{pos}^{(s)}, g_{dens}^{(s)}\} \in \mathbb{R}^{B \times 3 \times 1}$ to adaptively scale the ancestor, position, and density features; and 2) a feature generator that produces a dedicated loss-rate-aware feature $\mathbf{F}_p^{(s)} \in \mathbb{R}^{B \times 64 \times N_s}$.

The intermediate features from the parallel branches are first modulated by the gating factors $\mathbf{G}^{(s)}$ via element-wise multiplication. These modulated features are concatenated with $\mathbf{F}_p^{(s)}$, resulting in a fused tensor of shape $(B, 256, N_s)$. A final Conv1d layer reduces the dimension back to 64, which is then added to the sampled features via a residual connection to form the stage-specific latent features $\mathbf{y}^{(s)}$.

\subsection{Variational Hyperprior Network}
The latent features $\mathbf{y}$ are rearranged into a 2D representation $\mathbf{y}_{2D}^i \in \mathbb{R}^{B \times 64 \times H \times W}$ for entropy coding. The hyperprior network, detailed in Figure~\ref{fig:hyperprior_restoration}, follows a variational autoencoder structure \cite{[51]}.

The Hyper-encoder consists of three Conv2d layers. The first two layers employ a kernel size of 5 and a stride of 2, reducing the spatial resolution to $(H/2, W/2)$ and $(H/4, W/4)$, respectively. The channel dimension is adjusted from $64 \to 48 \to 48$. The final layer produces the hyper-latent $\mathbf{z}$. The Hyper-decoder mirrors this process using Deconv2d layers to restore the resolution. As shown in the Hyper-decoder section of Figure~\ref{fig:hyperprior_restoration}, the network branches into two parallel heads to predict the distribution parameters $\boldsymbol{\mu}$ and $\boldsymbol{\sigma}$. The scale head specifically includes a Softplus activation and a LowerBound operator to ensure the standard deviation remains positive and numerically stable.

\subsection{Latent Restoration: MGLR and DBR}
\begin{figure*}[h]
  \centering
  \includegraphics[width=\textwidth]{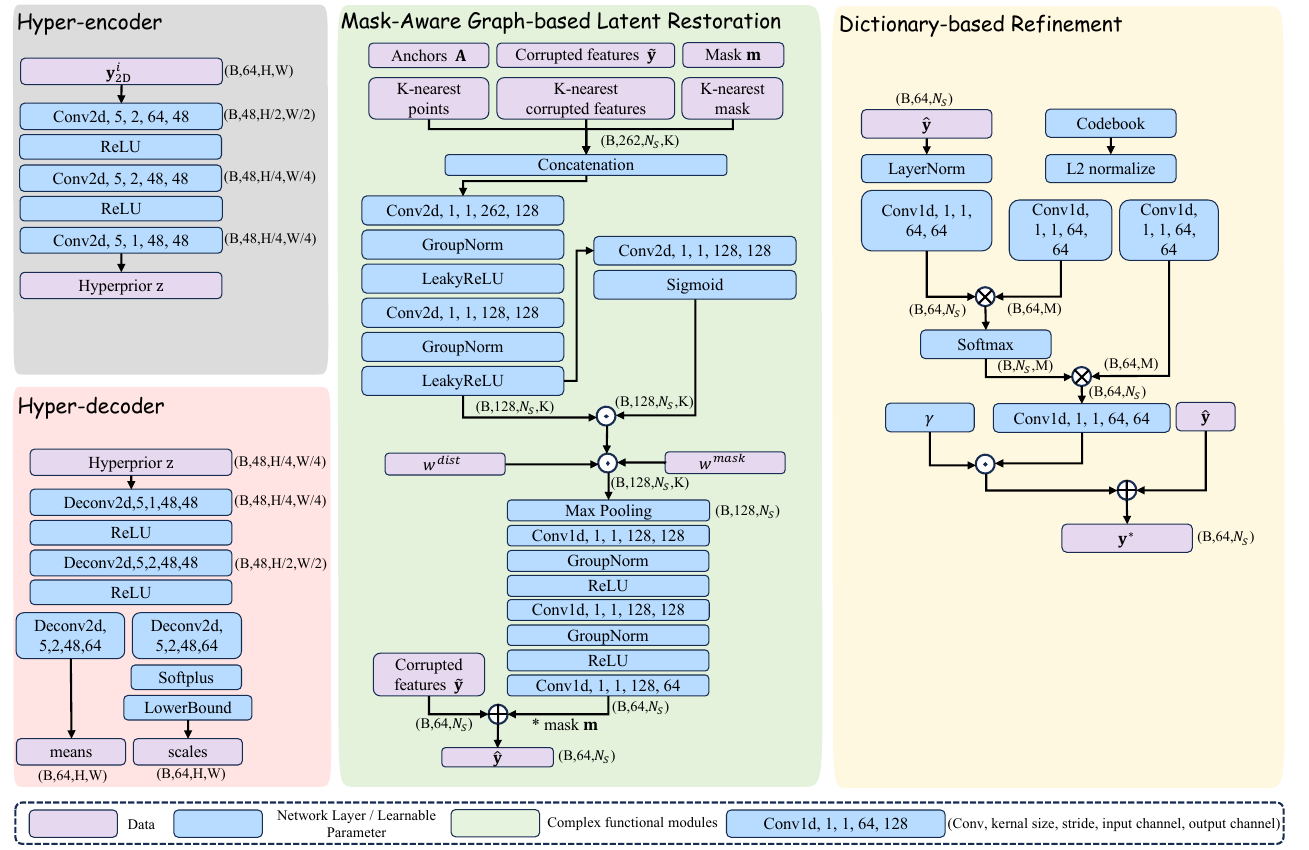}
  \caption{Detailed architecture of the Hyperprior network and the restoration pipeline. MGLR restores corrupted latents using graph-based neighborhood aggregation, followed by DBR for dictionary-based feature refinement.}
  \label{fig:hyperprior_restoration}
\end{figure*}

At the decoder, the corrupted latent features $\tilde{\mathbf{y}}$ are restored through a dual-stage pipeline illustrated in Figure~\ref{fig:hyperprior_restoration}. The first stage is the Mask-Aware Graph-based Latent Restoration (MGLR). It constructs a local graph where edge descriptors are generated by concatenating anchor coordinates $\mathbf{A}$, corrupted features $\tilde{\mathbf{y}}$, and the loss mask $\mathbf{m}$. For each neighbor in the $K$-nearest graph, an MLP computes an edge feature of shape $(B, 128, N_s, K)$. These features are then weighted by $w^{mask}$ and $w^{dist}$. A Max Pooling operation across the $K$ dimension aggregates the neighbor information to produce the restored features $\hat{\mathbf{y}}$.

The second stage is the Dictionary-based Refinement (DBR). This module employs a learnable codebook $\mathbf{D} \in \mathbb{R}^{128 \times 64}$ to regularize the features. The restored features $\hat{\mathbf{y}}$ undergo LayerNorm and are projected to generate queries, while keys and values are derived from the L2-normalized codebook. A cross-attention mechanism computes the similarity between $\hat{\mathbf{y}}$ and the prototype patterns in the codebook. The retrieved prior information is scaled by a learnable factor $\gamma$ and added back to $\hat{\mathbf{y}}$ via a residual connection, resulting in the final refined latent representation $\mathbf{y}^* \in \mathbb{R}^{B \times 64 \times N_s}$.

\section{Loss Functions and Evaluation Metrics}

This section provides detailed descriptions of the loss functions used for model optimization and the metrics employed for evaluation.

\subsection{Loss Function and Optimization}

The ResPCC framework is optimized end-to-end using a joint Rate-Distortion (R-D) objective function:
\begin{equation}
L = D + \lambda R,
\end{equation}
where $D$ represents the total reconstruction distortion, $R$ denotes the estimated bitrate, and $\lambda$ is the Lagrange multiplier.

\subsubsection{Distortion Loss}
The total distortion $D$ is a weighted sum of three components:
\begin{equation}
D = D_{cha} + \alpha D_{den} + \beta D_{card},
\end{equation}
where $\alpha$ and $\beta$ are hyperparameters balancing the respective terms.

\textbf{Chamfer Distance ($D_{cha}$):} To supervise geometry reconstruction and prevent error accumulation, we compute the symmetric point-to-point Chamfer Distance (CD) \cite{[31]} across all $S$ decoding stages. For the ground truth point cloud $\mathbf{X}_s$ and the reconstructed point cloud $\hat{\mathbf{X}}_s$ at stage $s$, the CD is defined as:
\begin{equation}
CD(\mathbf{X}_s, \hat{\mathbf{X}}_s) = \frac{1}{|\mathbf{X}_s|} \sum_{\mathbf{x}_i \in \mathbf{X}_s} \min_{\hat{\mathbf{x}}_j \in \hat{\mathbf{X}}_s} \|\mathbf{x}_i - \hat{\mathbf{x}}_j\|_2^2 + \frac{1}{|\hat{\mathbf{X}}_s|} \sum_{\hat{\mathbf{x}}_j \in \hat{\mathbf{X}}_s} \min_{\mathbf{x}_i \in \mathbf{X}_s} \|\hat{\mathbf{x}}_j - \mathbf{x}_i\|_2^2.
\end{equation}
The total geometric distortion is aggregated as:
\begin{equation}
D_{cha} = \sum_{s=0}^{S-1} CD(\mathbf{X}_s, \hat{\mathbf{X}}_s).
\end{equation}

\textbf{Cardinality Loss ($D_{card}$):} This term ensures global density consistency by penalizing the difference in total point counts at each stage:
\begin{equation}
D_{card} = \sum_{s=0}^{S-1} \big| |\mathbf{X}_s| - |\hat{\mathbf{X}}_s| \big|.
\end{equation}

\textbf{Density Loss ($D_{den}$):} To refine the local distribution, we introduce a density-aware term. For each upsampled point $\hat{\mathbf{x}}$ in the decoder, we identify its nearest counterpoint $\mathbf{x}$ in the encoder. Let $\mathcal{C}(\mathbf{x})$ and $\hat{\mathcal{C}}(\hat{\mathbf{x}})$ denote the collapsed and upsampled local point sets, respectively. The density loss is formulated as:
\begin{equation}
D_{den} = \sum_{s=0}^{S-1} \sum_{\hat{\mathbf{x}} \in \hat{\mathbf{X}}_{s+1}} \frac{ \big| |\mathcal{C}(\mathbf{x})| - |\hat{\mathcal{C}}(\hat{\mathbf{x}})| \big| + \gamma \Delta\mu + \gamma \Delta\sigma }{ |\hat{\mathbf{X}}_{s+1}| },
\end{equation}
where $\Delta\mu = |\overline{\mathcal{C}(\mathbf{x})} - \overline{\hat{\mathcal{C}}(\hat{\mathbf{x}})}|$ represents the difference in mean distances to the center points, and $\Delta\sigma = |std(\mathcal{C}(\mathbf{x})) - std(\hat{\mathcal{C}}(\hat{\mathbf{x}}))|$ measures the difference in the standard deviation of these distances. Both geometric distribution terms share the same weighting coefficient $\gamma$.

\subsubsection{Rate Loss}
We replace the non-differentiable quantization step with additive uniform noise $\mathcal{U}(-0.5, 0.5)$ during training \cite{[1]}. The rate loss $R$ is estimated as the number of bits required to encode the latent representation and hyperprior based on the predicted probability distributions. During inference, features are properly quantized for arithmetic coding.

\subsection{Evaluation Metrics}

We evaluate the geometric reconstruction quality using the standard metrics defined in the MPEG point cloud compression common test conditions (CTC). Specifically, we report the point-to-point PSNR (D1) and point-to-plane PSNR (D2) \cite{[54]}.

\subsubsection{Mean Square Error (MSE) Calculation}
The PSNR values are derived from the Mean Square Error (MSE) between the ground truth point cloud $\mathbf{X}$ and the reconstructed point cloud $\hat{\mathbf{X}}$.

\textbf{Point-to-point MSE ($e_{D1}$):} For each point $\mathbf{x}_i$ in the ground truth set, the distance to its nearest neighbor $\hat{\mathbf{x}}_j$ in the reconstructed set is computed. The symmetric $e_{D1}$ is defined as:
\begin{equation}
e_{D1} = \frac{1}{|\mathbf{X}|} \sum_{\mathbf{x}_i \in \mathbf{X}} \min_{\hat{\mathbf{x}}_j \in \hat{\mathbf{X}}} \|\mathbf{x}_i - \hat{\mathbf{x}}_j\|_2^2.
\end{equation}

\textbf{Point-to-plane MSE ($e_{D2}$):} This metric measures the projected distance along the surface normal. Let $\mathbf{n}_i$ be the normal vector of point $\mathbf{x}_i$. The $e_{D2}$ error is formulated as:
\begin{equation}
e_{D2} = \frac{1}{|\mathbf{X}|} \sum_{\mathbf{x}_i \in \mathbf{X}} \left( \min_{\hat{\mathbf{x}}_j \in \hat{\mathbf{X}}} (\mathbf{x}_i - \hat{\mathbf{x}}_j) \cdot \mathbf{n}_i \right)^2.
\end{equation}
The D2 metric provides a better reflection of the surface reconstruction accuracy and is less sensitive to sampling density variations.

\subsubsection{PSNR Formulation}
The final PSNR (in dB) is calculated using the peak value $p$, which corresponds to the maximum possible distance in the coordinate space (the diagonal of the bounding box). The formula is:
\begin{equation}
PSNR = 10 \log_{10} \left( \frac{p^2}{MSE} \right),
\end{equation}
where $MSE$ is either $e_{D1}$ or $e_{D2}$ for the respective PSNR types.

\section{Implementation and Training Details}

In addition to the primary configurations described in the main paper, this section provides specific hyperparameter settings to ensure the reproducibility of our experiments. The entire ResPCC framework is implemented using PyTorch \cite{[58]} and trained on a single NVIDIA RTX 3090 GPU.

\subsection{Optimizer and Scheduler Settings}
We optimize the network parameters using the Adam optimizer \cite{[59]}. The initial learning rate is set to $1 \times 10^{-3}$ for the backbone codec and the entropy model. The optimizer momentum parameters $\beta_1$ and $\beta_2$ are configured as 0.9 and 0.999, respectively. 

To facilitate stable optimization, we employ a StepLR learning rate scheduler with a decay factor $\gamma$ of 0.5. The decay interval is adjusted based on the scale of the dataset. For the ShapeNet \cite{[47]} dataset, the learning rate decays every 10 epochs. For the SemanticKITTI \cite{[48]} dataset, we increase the decay step to 15 epochs to accommodate its larger data volume.

\subsection{Loss Function Coefficients}
The weighting coefficients for the distortion terms are consistent across both datasets to ensure a uniform optimization objective. For the total distortion loss, the coefficient for the multi-stage Chamfer Distance ($D_{cha}$) is set to 0.5. The cardinality loss ($D_{card}$) is weighted by $\beta = 5 \times 10^{-7}$, and the density-aware loss ($D_{den}$) is assigned a weight of $\alpha = 1 \times 10^{-4}$. Within the density loss formulation, the coefficients for both the mean distance difference ($\Delta\mu$) and the standard deviation difference ($\Delta\sigma$) are set to $\gamma = 50$.

\subsection{Architectural Hyperparameters}
The internal sub-modules of ResPCC follow a consistent set of structural hyperparameters across different datasets. The number of nearest neighbors $K$ is fixed at 8 for both the geometric feature extraction in the downsampling layers and the graph construction in the MGLR module. The hidden dimension for the MLPs within the CALM, MGLR, and DBR modules is set to 128. For the variational hyperprior network, we use 48 channels for the hyper-latent representation and intermediate convolutional layers. Furthermore, all group normalization layers throughout the network are configured with 4 groups to balance batch-level and channel-level statistics.

\subsection{Evaluation Constants}
The peak values for PSNR calculation are determined by the maximum spatial range of each dataset to provide a normalized comparison. For the ShapeNet dataset, the peak value is set to 3.4641. For the SemanticKITTI dataset, we use a peak value of 152.0625. These constants are applied consistently across all D1-PSNR and D2-PSNR evaluations. All baseline methods are evaluated using the same peak values and hardware environment to ensure a fair comparison under various packet loss conditions.

\section{Additional Ablation Studies}

We evaluate the impact of different packet loss rate (PLR) sampling strategies during training. We compare our proposed strategy with a baseline strategy that uses uniform stochastic sampling. In the uniform strategy, the PLR is sampled from a uniform distribution between 0\% and 50\%. Our strategy sets the PLR to zero with a probability of 50\% and samples from a Beta distribution ($\alpha=1.1, \beta=2$) otherwise. For a fair comparison, we keep the network architecture identical and exclude the DBR module. We conduct the evaluation on a sampled subset of 55 ShapeNet test instances.

Table~\ref{tab:ablation_strategy_combined} shows the experimental results. Our strategy consistently achieves higher PSNR values and lower bitrates under common packet-free and low-loss conditions. For example, at 0\% PLR, our strategy reaches 70.62 dB (D2-PSNR) with 2.390 Bpp, while the uniform strategy only reaches 69.50 dB with 2.454 Bpp. Although the baseline uniform sampling shows slightly better PSNR at the extreme 30\% loss rate, its performance is restricted in lower loss scenarios. This aligns with the findings in \cite{[42]}, where a significant drop in quality is observed under low loss rates while the quality improvement under high loss rates is only marginal. Our strategy provides a better balance for practical network environments where low-loss scenarios are more frequent.

\begin{table}[h]
\caption{Comparison of different PLR sampling strategies on ShapeNet (without DBR). Values in bold indicate better performance.}
\label{tab:ablation_strategy_combined}
\centering
\small
\begin{tabular}{llcccc}
\toprule
PLR & Strategy & Bpp $\downarrow$ & Chamfer Dist. $\downarrow$ & D2-PSNR $\uparrow$ & D1-PSNR $\uparrow$ \\
\midrule
0\%  & Ours & \textbf{2.390} & \textbf{9.48E-06} & \textbf{70.62} & \textbf{64.56} \\
     & Uniform  & 2.454 & 1.05E-05 & 69.50 & 64.10 \\
\midrule
5\%  & Ours & \textbf{2.417} & \textbf{9.72E-06} & \textbf{70.21} & \textbf{64.47} \\
     & Uniform  & 2.477 & 1.06E-05 & 69.35 & 64.07 \\
\midrule
10\% & Ours & \textbf{2.444} & \textbf{9.99E-06} & \textbf{69.77} & \textbf{64.37} \\
     & Uniform  & 2.504 & 1.07E-05 & 69.19 & 64.04 \\
\midrule
20\% & Ours & \textbf{2.505} & \textbf{1.07E-05} & \textbf{68.79} & \textbf{64.11} \\
     & Uniform  & 2.568 & 1.10E-05 & 68.75 & 63.92 \\
\midrule
30\% & Ours & \textbf{2.582} & 1.18E-05 & 67.67 & \textbf{63.75} \\
     & Uniform  & 2.646 & \textbf{1.15E-05} & \textbf{68.13} & 63.74 \\
\bottomrule
\end{tabular}
\end{table}

\section{More Qualitative Results}

\begin{figure*}[ht]
  \centering
  \includegraphics[width=0.92\textwidth]{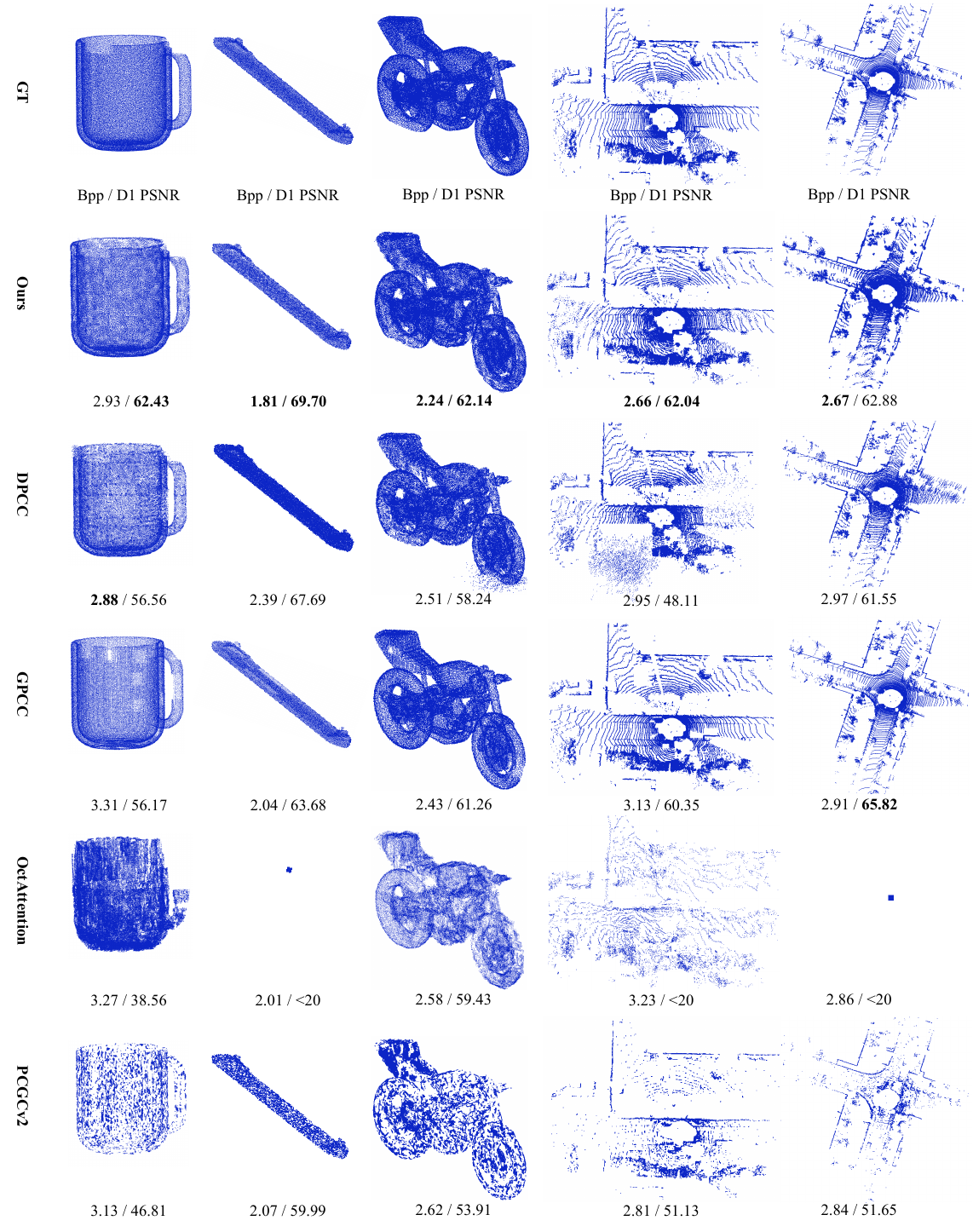} 
  \caption{Visual comparison at 5\% PLR. ResPCC and baselines show relatively stable results, though our method achieves higher PSNR and better detail preservation.}
  \label{fig:plr5}
\end{figure*}

\begin{figure*}[ht]
  \centering
  \includegraphics[width=0.92\textwidth]{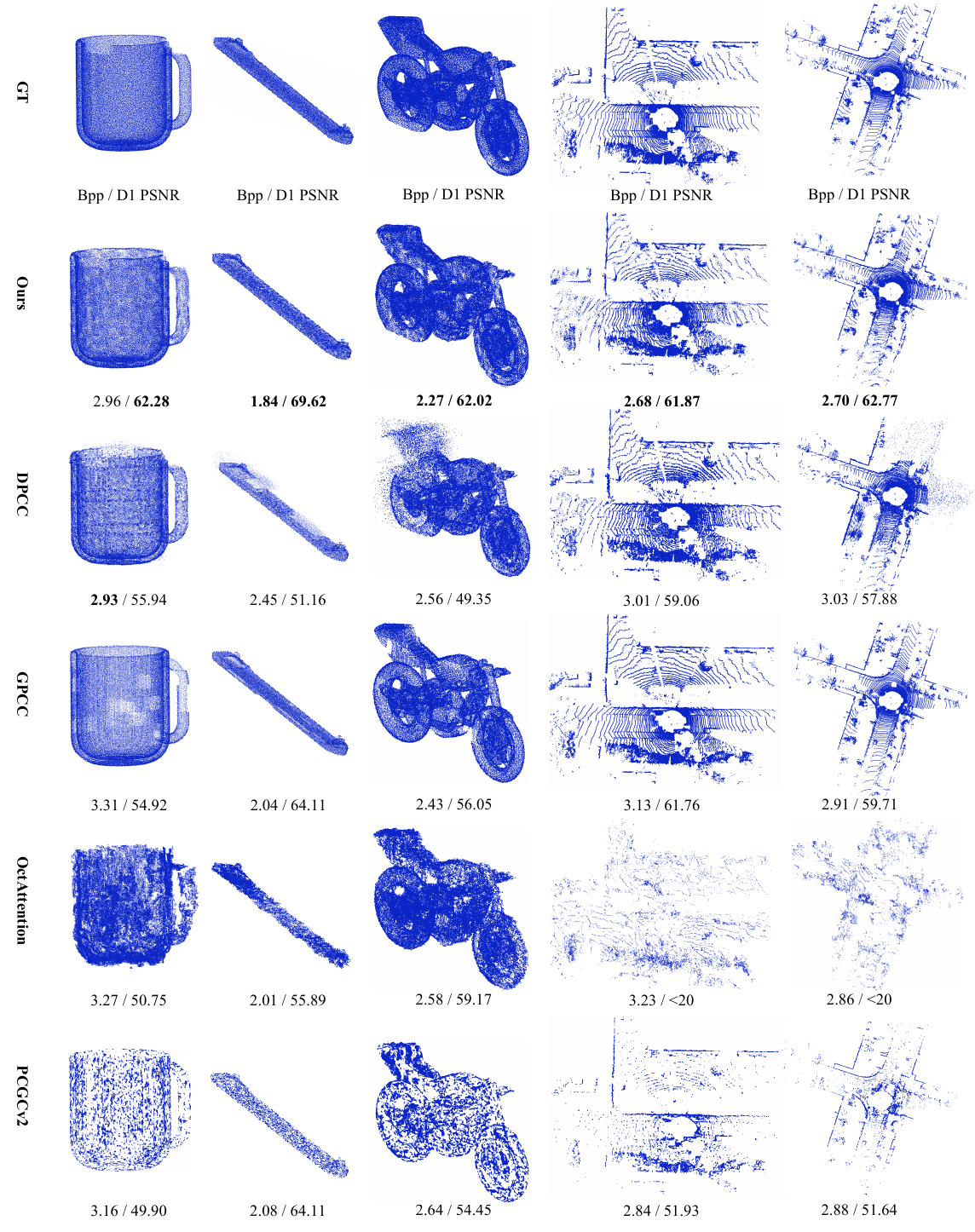} 
  \caption{Visual comparison at 10\% PLR. Noticeable sparsity starts appearing in PCGCv2 and OctAttention, while ResPCC remains consistent with the ground truth.}
  \label{fig:plr10}
\end{figure*}

\begin{figure*}[ht]
  \centering
  \includegraphics[width=0.92\textwidth]{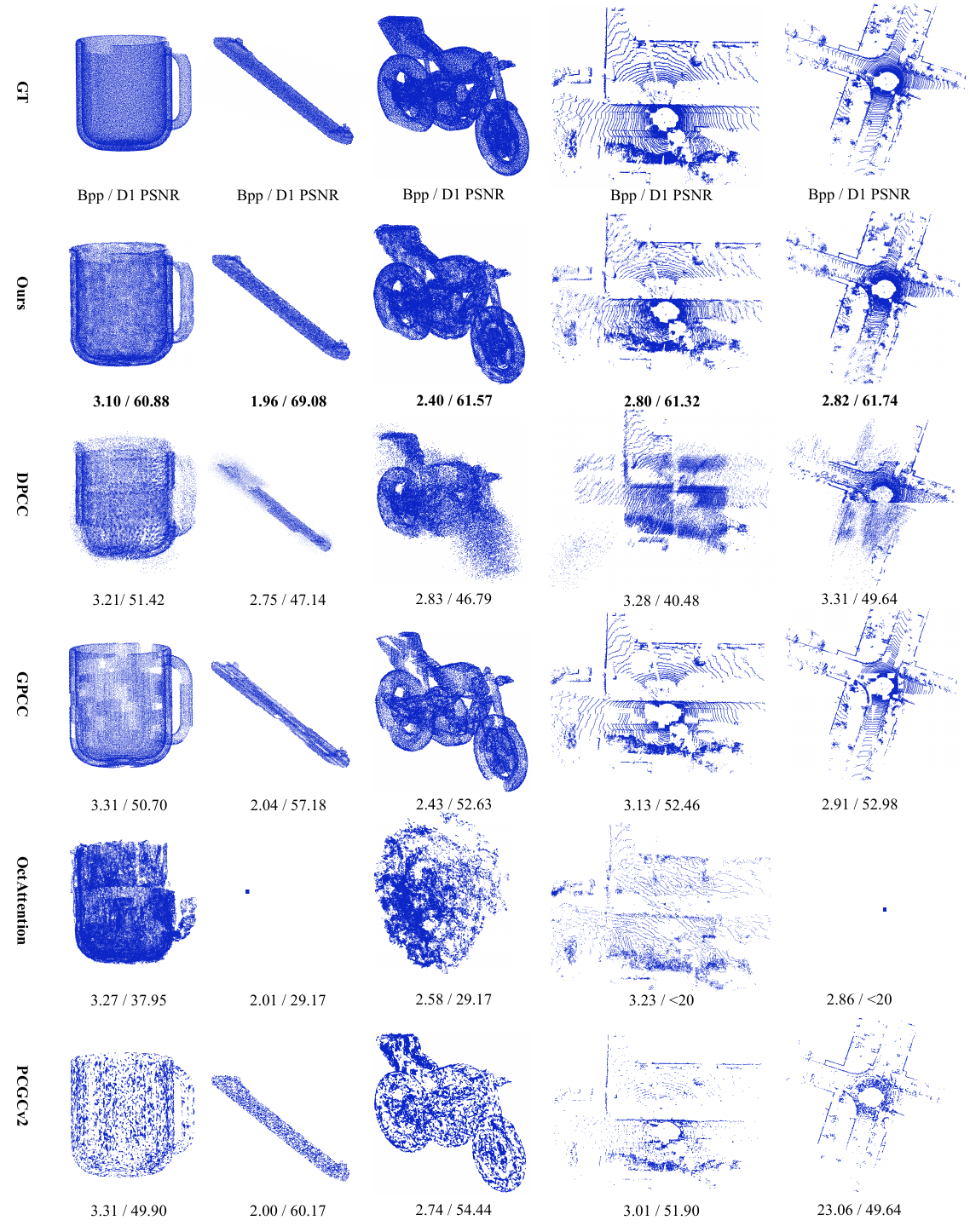} 
  \caption{Visual comparison at 20\% PLR. OctAttention exhibits extreme instability, whereas ResPCC effectively restores the geometry.}
  \label{fig:plr20}
\end{figure*}

\begin{figure*}[ht]
  \centering
  \includegraphics[width=0.92\textwidth]{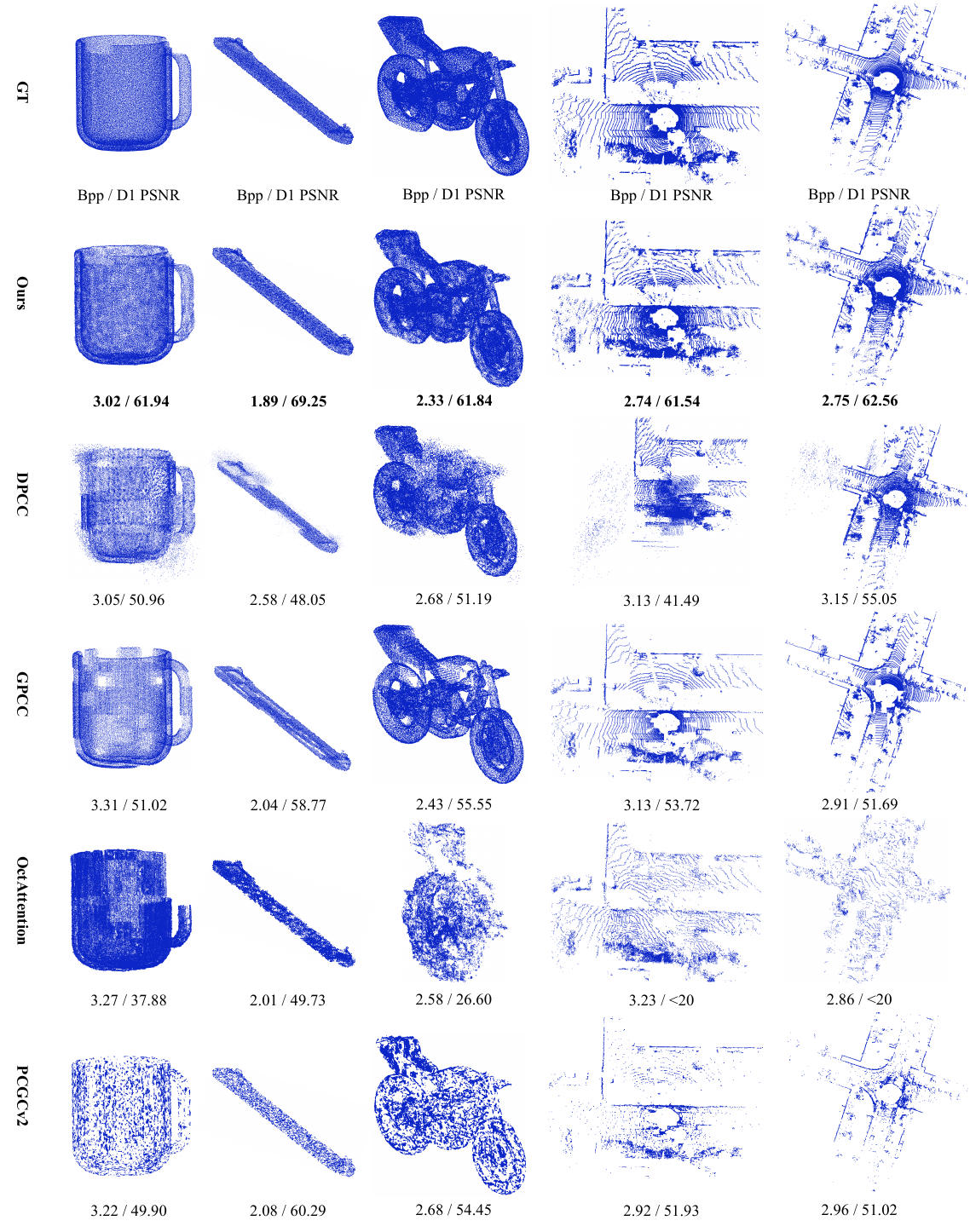} 
  \caption{Visual comparison at 30\% PLR. Under severe loss, ResPCC is the only method that consistently preserves a recognizable and complete object structure.}
  \label{fig:plr30}
\end{figure*}

We provide extensive qualitative comparisons on the ShapeNet and SemanticKITTI datasets to further evaluate the reconstruction fidelity of ResPCC under various transmission conditions. Figures~\ref{fig:plr5} to \ref{fig:plr30} illustrate the reconstructed point clouds under PLR of 5\%, 10\%, 20\%, and 30\%, respectively. As shown in these results, ResPCC consistently maintains high geometric integrity and clear object boundaries across diverse categories. In contrast, the performance of baselines degrades significantly as the loss rate increases. 

A critical observation is the extreme instability of octree-based methods like OctAttention \cite{[34]}. While the main paper included relatively successful cases, these supplementary figures highlight its high performance variance. The reconstruction quality of OctAttention depends heavily on the octree depth and spatial position where packet loss occurs. In several instances under 20\% and 30\% PLR (e.g., the airplane and motorbike samples), OctAttention suffers from total topological collapse or the loss of major components, resulting in PSNR values dropping below 20 dB. Other baselines also exhibit distinct artifacts. MPEG G-PCC shows blocky regional holes due to its spatial slicing, while DPCC \cite{[17]} suffers from global coordinate drift and blurred surfaces. By leveraging the adaptive modulation of CALM, the error dispersion mechanism of SCI, and the restoration capabilities of MGLR and DBR, ResPCC effectively mitigates these issues, providing a reliable and stable solution for 3D point cloud transmission over lossy networks.

\balance

\end{document}